\documentclass{rspublicchanged}

\begin{document}

\newcommand{\ket}[1]{\ensuremath{|#1 \rangle}}
\newcommand{\bra}[1]{\ensuremath{\langle #1|}}
\newcommand{\braket}[2]{\ensuremath{\langle #1|#2 \rangle}}
\newcommand{\ro}[1]{\ensuremath{|#1 \rangle \langle #1|}}

\newcommand{\real}{\ensuremath{\mathrm{Re}}}
\newcommand{\trace}{\ensuremath{\textsf{Tr}}}

\newcommand{\id}{\ensuremath{\mathsf{1}}}
\newcommand{\R}{\ensuremath{{\sf R\hspace*{-0.9ex}\rule{0.15ex}
{1.5ex}\hspace*{0.9ex}}}}
\newcommand{\N}{\ensuremath{{\sf N\hspace*{-1.0ex}\rule{0.15ex}
{1.3ex}\hspace*{1.0ex}}}}
\newcommand{\Q}{\ensuremath{{\sf Q\hspace*{-1.1ex}\rule{0.15ex}
{1.5ex}\hspace*{1.1ex}}}}
\newcommand{\C}{\ensuremath{{\sf C\hspace*{-0.9ex}\rule{0.15ex}
{1.3ex}\hspace*{0.9ex}}}}

\newcommand{\h}[1]{\ensuremath{\mathcal{H}_{#1}}}

\newcommand{\me}{\ensuremath{\mathrm{e}}}
\newcommand{\mi}{\ensuremath{\mathrm{i}}}

\newcommand{\de}{\ensuremath{\mathrm{d}}}
\newcommand{\dd}[2]{\ensuremath{\frac{\mathrm{d}#1}{\mathrm{d}#2}}}
\newcommand{\ddd}[2]{\ensuremath{\frac{\mathrm{d}^2#1}{\mathrm{d}#2^2}}}

\newcommand{\ot}[2]{\ensuremath{\left( \begin{array}{c} #1 \\ #2
\end{array} \right)}}
\newcommand{\oth}[3]{\ensuremath{\left( \begin{array}{c} #1 \\ #2 \\ #3
\end{array} \right)}}
\newcommand{\twtw}[4]{\ensuremath{\left( \begin{array}{cc} #1 & #2 \\
#3 & #4 \end{array} \right)}}
\newcommand{\thth}[9]{\ensuremath{\left( \begin{array}{ccc} #1 & #2 & #3
\\ #4 & #5 & #6 \\ #7 & #8 & #9 \end{array} \right)}}

\newcommand{\expp}[1]{\ensuremath{\me^{\mi\hat{H}#1}}}
\newcommand{\expm}[1]{\ensuremath{\me^{-\mi\hat{H}#1}}}

\title{Quantum Computation and Many Worlds}
\author[C Hewitt-Horsman]{Clare Hewitt-Horsman}
\label{firstpage}
\maketitle

\begin{abstract}{Quantum computation, information processing, many worlds}
 An Everett (`Many Worlds') interpretation of quantum mechanics due to Saunders and Zurek is presented in detail.
 This is used to give a
 physical description of the process of a quantum computation. Objections to such an understanding are discussed.
\end{abstract}

\section*{Introduction}

In quantum computation at the moment we do not have a physical picture of what is actually going on in a
computation, as this is dependent on the interpretation that one has of quantum mechanics. In this paper we will
look at what is going on in a general quantum computational process, from the point of view of the Everett (`Many
Worlds') interpretation of quantum mechanics. It is sometimes argued that quantum computation is to be regarded as
proof of an Everett-style interpretation (Deutsch 1997; Wallace, in preparation).
This is \emph{not} the aim of the current paper.\\

In this paper we will not deal with the arguments for or against the particular Everett interpretation used here (a
combination of the interpretations of Saunders (1995, 1996) and Zurek (1991, 1993, 2001)\footnote{There is also a
lot of similarity with the Everett-style theories used by Gell-Mann \& Hartle (1990).}, recently expanded and
explicated by Wallace (2001$a$,$b$)), nor against Everett interpretations in general. However there are some
arguments that appear to object to using Everett interpretations in quantum computation at all, and these we shall
address, most notably that given by Steane (2000). The problem with such arguments is that they often seem to be
attacking an Everett theory that does not really exist: there does not seem to be a real understanding of exactly
what an Everett theory says is going on in a quantum
computation. The purpose of the present paper is to provide such an understanding in detail.\\

The structure of the paper is as follows.\\
\noindent  In {\bfseries section 1} is presented the Everett theory that we will be using in the rest of the paper. \\
\noindent In {\bfseries section 2} our Everett theory is used to give a physical picture of quantum computation.\\
\noindent In the light of the previous sections, in {\bfseries section 3} we look at some objections often raised
against using an Everett theory to describe a quantum computation process.

\section{The Everett-Saunders-Zurek Interpretation}

The interpretation of quantum mechanics has been a problem since the very beginnings of the theory. It is not the
intention here to enter into a detailed discussion of the problem and the various proposed solutions; such a study
can be found in many places, for example Davies \& Brown (1986). We will, however, briefly outline the main
difficulty (the `measurement problem') before going on to describe the proposed solution that is of relevance here.

\subsection{The Measurement Problem in Brief}

As we all know, quantum mechanics predicts undetermined states for microscopic objects most of the time: for
example, in an interferometer the photon path is indeterminate between the two arms of the apparatus.
We deal with such states all the time, and are seemingly happy with them for the unobservable realm.\\

Such happiness is destroyed when we consider an experiment (such as the infamous Schr\"{o}dinger's Cat set-up)
where macroscopic outcomes are made dependent on microscopic states. We are then faced with an `amplification of
indeterminism' up to the macro-realm: the state of the system+cat is something like
\begin{equation} \ket{0} \otimes \ket{\mathrm{cat \ dead}} + \ket{1} \otimes \ket{\mathrm{cat \ alive}}
\label{cat} \end{equation}

 This is the measurement problem: how do we
reconcile the fact that quantum mechanics predicts macroscopic indeterminism with the fact that we observe a
definite macro-realm in everyday life?

\subsection{ESZ}\label{descript}

Almost all proposed solutions to the measurement problem start from this assumption: that a superposition of states
such as (\ref{cat}) cannot describe macroscopic physical reality. In one way or another all terms bar one are made
to vanish, leaving the `actual' physical world.\\

The exception to this way of solving the problem was first proposed by Everett (1957). His interpretation has since
been modified and improved, but the central posit remains the same: that physical reality at all levels is
described by states such as (\ref{cat}), and each term in such a superposition is equally
actualized. \\

Dispute over what this actually \emph{means} gives rise to the myriad of different Everettian interpretations that
we have (Many Worlds, Many Minds, etc. etc. etc.). One thing we can say about all Everett interpretations is that
they include at some level and to some degree a multiplicity of what we would commonly refer to as the `macroscopic
world': each term in the superposition is actual, the cat is \emph{both} dead \emph{and} alive. In general this
gives rise to the `incredulous stare' argument, arguing against such a huge increase in ontological commitment.
However, as has always been the case in physics, extra ontology is welcomed into the theory when it does
explanatory work (for example the extra ontology required for the huge universe of modern
cosmology over previous theories).\\

There are two pressing problems at this point for the Everettian\footnote{Actually there are three, but the third
(the problem of probability --- see eg Saunders (1998)) lies well outside the scope of this paper.}. Firstly, the
measurement problem itself: if all terms are real, why do we only see one? Secondly, there is the logical
problem: how can anything be in two states at once?\\
Taking the second problem first, we note that we do not get a logical contradiction when something is in two
different states with respect to an external parameter. For example, I am both sitting at my desk typing and
standing by the window drinking tea, with respect to the parameter time: \emph{now} I am at my desk, but
(hopefully) \emph{soon} I will be drinking tea. The parameter in Everett theories with respect to which cats, etc.,
have their properties is variously called the world, branch, universe, or macrorealm (see below for a note on
terminology). The idea (very very basically) is that in one world (branch, etc.) the cat is dead, and in another it
is alive. Extending this to include experimenters we get the answer to our first question: in one world the
experimenter sees the cat dead, in another she sees it alive.\\

We now have the problem of making these rather vague ideas concrete. As noted above, the differing ways of doing
this give rise to different Everett-style interpretations. We shall now turn to a specific theory (chosen as the
best of the Everett-style theories on offer), an amalgam of the ideas of Everett (1957), Saunders (1995, 1996) and
Zurek (1991, 1993, 2001) (and the expansion of these by Wallace (2001$a$,$b$)), which we will call the
Everett-Saunders-Zurek interpretation, hereafter ESZ.\\

The main ideas of the interpretation are the following. Physical reality is represented by the state $\ket{\Psi}$:
the `state of the universe'. Within this main structure we can identify substructures that behave like what we
would intuitively call a single universe (the sort of world we see around us). There are many such substructures,
which we identify as single universes with different histories. The identification of these substructures is not
fundamental or precise - rather, they are defined for all practical purposes (FAPP), using criteria such as
distinguishability and stability over time. Decoherence plays an important role in identifying macroscopic worlds,
but (as will be important when we go on to talk about quantum computation) worlds can be identified when the system
is fully coherent. A further concept is where distinguishable structures are not stable over time: we can still
identify worlds structures, but only at a point in time.

A note on terminology before we proceed further.

\begin{description}
\item[Multiverse] The totality of physical reality, described by the `universal state'. There is no larger system
than this: this is the main structure.

\item[Branch] A substructure of the multiverse that we would intuitively call a `universe': a distinct macrorealm
with a definite (or quasi-definite) history evolving stably over time. This is the level of description that we
need to recover in order to  solve the measurement problem.

\item[World] A world is similar to a branch, only the structures thus identified are not stable over long
timescales. We can identify them within the main state over relevant timescales (that are long enough for the
structures to be useful to us), but they do not evolve stably and keep their structure for longer times.

\end{description}

In these terms, the claim of ESZ is that physical reality is the multiverse, and within that there are branches
that we as observers inhabit. States such as (\ref{cat}) describe the multiverse, with each term in the
superposition being a branch, that being what an experimenter sees. As well as these long-lasting branches there
are also identifiable worlds within the main structure, at many different levels -- within a branch we may have a
structure within which many worlds may be identifiable for a certain length of time.\\

There are now further questions that must be answered: how are the branches and worlds defined, and how does their
stability arise? \\

The main points here are that branches and worlds keep their structure for long enough to be useful, and that they
evolve essentially independently of other structures. Because these structures are not fundamental ontology it does
not matter that they are defined thus roughly (and indeed have no precise definition) -- what matters is that they
stay recognizable for long enough for us to frame theories about their behaviour (see \S\ref{fs} below for
additional comments on this).\\

For branches, stability and FAPP independence comes from a definite mechanism: decoherence. Decoherence is the
lynchpin of ESZ's response to the measurement problem, allowing the stable evolution of definite substructures of
universes within the universal state. I will not here go through all the mechanics of the decoherence process (this
may be found in many places, for example Joos \& Zeh (1985) and Omn\'{e}s (1994)), but merely state the relevant points.\\

Decoherence occurs when a system becomes entangled with a larger environment. If we then look at the behaviour of
the system alone then we have to trace out the environment - which, as is well known, leads to the loss of the
interference terms between states of the decoherence basis. Thus, at any given instant, we can identify a
multiplicity of non-interfering substructures, which are (in the above terminology) worlds. Furthermore this lack
of interference persists over time if decoherence is ongoing: that is, individual substructures (elements of the
decoherence basis) evolve virtually independently, with minimal interference with other such substructures.

\subsection{FAPP Structures}\label{fs}

The foregoing, or similar has been the claim of Everett theories ever since decoherence came on the scene; what
then is different about ESZ? The difference comes about in the way in which this specification of the branches (and
worlds) is handled. In, for example, a naive many worlds interpretations (eg Deutsch 1985) one cannot find the
preferred basis to specify the branches using decoherence as the branch structure is absolute and decoherence, as
is well known, is only an approximate process: we get a specification of branches for all practical purposes
(FAPP), but not in any absolute sense. It is a common assertion in the literature on the preferred basis problem
that the preferred basis must be defined absolutely, explicitly and precisely, and that therefore the failure to
give such a definition (and indeed the impossibility of doing so) is a terminal
problem for an Everett interpretation.\\

The difference with ESZ is that the branch and worlds structure is \emph{not} absolute, and so no such explicit or
precise rule is required\footnote{This vital understanding is found in Wallace (2001$b$), from which the material
for this section is taken}. The key to understanding how this works is to move away from the idea, implicit in much
of the literature, that the measurement problem must be solved at the level of the basic theory: that is, that we
must recover a single macrorealm (or the appearance of one) in the fundamental posits of our theory. ESZ does
something different, by defining the worlds, in Wallace's phrase, as `higher order ontology'. The structures of the
worlds and branches are not precisely defined, but FAPP they are a useful way to look at the overall state.
Furthermore, we as observers are some such structures, and so we must look at the evolution of these structures and
the rest of the state relative to them in order to recover predictions about the branch we live in from quantum
mechanics -- ultimately, to answer the measurement problem.\\

Physics (and indeed science in general) is no stranger to the idea of using approximately-defined structures. In
everyday life we deal with them all the time. For example, we can go to the beach and (if we are in a suitably
meditative mood) count the waves as they come in. If we are feeling more energetic then we can paddle out and use
one particular wave to surf on. The waves exists as real entities (I can count them and surf on them), they persist
over time as distinct structures (we can follow them as they come into shore and break), and if I surfed on one
then I would talk about \emph{the} wave I caught.

Waves are, however, not precisely defined: where does this wave end and that one begin? Where does this one end and
the sea begin? Different water molecules comprise it at different points in its history -- given this, how is the
wave defined? We cannot find any method that will tell us absolutely when a given molecule is part of the wave or
not, and this is not merely a \emph{technical} impossibility: there is simply no fact of the matter about when a
wave ends and the sea begins. We can use rough rules of thumb, but at the fine-grained level there is no precise
fact to find.\\

We thus see that there are many objects that we would unhesitatingly call real that we nevertheless cannot define
absolutely and objectively. Such entities are part of our higher order ontology, not `written in' directly in the
fundamental laws of the theory, but nevertheless present and real and explanatorily useful.

It is at such a level that the ESZ concept of a world or branch operates. It is not an entity written into the
fundamental laws of the interpretation: in fact, what ESZ does is (merely?) explain how, from quantum mechanics
alone (decoherence is a straight consequence of the basic laws), the structures can emerge that describe the world
that we see around us everyday\footnote{This is in fact one of the great strengths of ESZ as an interpretation:
there are no mathematics added to standard quantum mechanics (a strength particularly for those physicists who do
not wish a theory to be changed for conceptual or philosophical reasons); it is truly an \emph{interpretation}.}.
These structures are not (and cannot be) absolutely defined, but this is no argument against their reality.

\section{Quantum Computation}

We now come to the main question of this paper: what physical picture does ESZ give us of a quantum computational
process in particular? \\

There is of course no one particular process that \emph{is} quantum computation: there are many different
algorithms, quite apart from their physical instantiation. However we will find that ESZ tells us things about how
information is processed in these algorithms \emph{in general}. Nevertheless, in working through this, it will be
useful to have a concrete example before us. For simplicity we will chose one of the simplest algorithms for our
example --- the Deutsch algorithm.

\subsection{The Deutsch Algorithm}

Suppose we have a function $f(x)$ whose domain and range are both $\{ 0,1\}$. We want to find out whether
$f(0)=f(1)$ or not. Classically, there is no quicker way to do this than evaluating both $f(0)$ and $f(1)$ and
comparing them. However there is a quantum algorithm that can do it with only \emph{one} evaluation --- this is the
Deutsch Algorithm.\\

\noindent To start the algorithm we take a two qubit register in the state
$$ \ket{01} = \ket{0}_a \otimes \ket{1}_b$$

\noindent We then perform two Hadamard transformations (one on each qubit), so that
\begin{equation} \ket{01} \rightarrow (\ket{0} + \ket{1}) (\ket{0} - \ket{1}) = \ket{\psi}\label{psi}\end{equation}

\noindent Let us now define the transformation $U_f$ as
$$ U_f : \ \ \ket{x,y} \rightarrow \ket {x, y\oplus f(x)}$$

\noindent where f(x) is our original function, and `$\oplus$' stands for addition modulo 2. For the state
$\ket{x}(\ket{0} - \ket{1})$ there are two options when acted on by $U_f$:
$$ \ket{x}(\ket{0} - \ket{1}) \rightarrow \left\{ \begin{array}{c} \ket{x}(\ket{0} - \ket{1}) \ \ \ \ f(x) = 0 \\
\ket{x}(\ket{1} - \ket{0}) \ \ \ \ f(x) = 1 \end{array} \right. $$

\noindent That is,
$$ \ket{x}(\ket{0} - \ket{1}) \rightarrow (-1)^{f(x)} \ket{x} (\ket{0} - \ket{1})$$

\noindent We therefore see that the action of $U_f$ on the state given by (\ref{psi}) is
\begin{eqnarray} U_f\ket{\psi} & = & (-1)^{f(0)} \ket{0}(\ket{0} - \ket{1}) + (-1)^{f(1)} \ket{1}(\ket{0}-\ket{1}) \label{correlate} \\
{} & = & \left\{ \begin{array}{c} \pm (\ket{0} + \ket{1})(\ket{0} - \ket{1}) \ \ \ \ f(0)=f(1) \\
\pm (\ket{0} - \ket{1})(\ket{0} - \ket{1}) \ \ \ \ f(0) \neq f(1)
\end{array} \right. \label{manip} \end{eqnarray}

\noindent We now act on the first qubit by another Hadamard transformation, so we are left with the state of our
two qubits as
\begin{eqnarray*} \pm \ket{0} (\ket{0} + \ket{1}) \ \ \ \ f(0)=f(1) \\
\pm \ket{1} (\ket{0} + \ket{1}) \ \ \ \ f(0) \neq f(1)
\end{eqnarray*}

\noindent If we now measure the first qubit then we will find out with probability 1 whether $f(0)=f(1)$ or not (if
we measure 0 then it is, if 1 then it is not) --- with one measurement.

\subsection{General Algorithms}

Following Wallace (in preparation), there are three main stages to the Deutsch algorithm that are common to all
quantum algorithms so far discovered that give a speed-up over their classical counterparts\footnote{We must note,
however, that there are still only a very few quantum algorithms that have been discovered, and that generalising
from such a small number is always a potentially risky business.}. They are the Hadamard transformations at the
beginning and end of the manipulation,
and the manipulation of the state itself.\\

\subsubsection{Hadamard 1}

In general, algorithms start by having two registers, the first of which is the input register (in the Deutsch
algorithm this is a single qubit, qubit a, but in general it will be $n$ qubits long). The register starts in the
state $\ket{0}^{\otimes n}$ (that is, $\ket{000...0}$ with $n$ lots of 0's) and then has $n$ Hadamard
transformations acted on it, resulting in a superposition of all its possible states,
$$ \frac{1}{\sqrt{2^n}} \sum_\alpha \ket{\alpha}$$

\noindent where $\alpha$ ranges over all these states. In our example above this is the action of step (\ref{psi})
on qubit a (note that the Hadamard performed on qubit b is {\bfseries not} an example of this general step),
resulting in the qubit going from state $\ket{0}$ to state $\ket{0} + \ket{1}$.

\subsubsection{Manipulation}

The common feature of the state manipulation is that states in the second register (above, qubit b) become
correlated with states of the first, and that the correlated states are a constant function of their states in the
first register. That is, that the state of the two registers evolves as
\begin{equation} \left( \frac{1}{\sqrt{2^n}} \sum_\alpha \ket{\alpha}\right) \otimes \ket{y} \rightarrow \frac{1}{\sqrt{2^n}} \sum_\alpha \ket{\alpha} \otimes \ket{g(\alpha)} \label{many}\end{equation}

\noindent where $\ket{y}$ is the initial state of the second register.\\

In the Deutsch algorithm this is shown in (\ref{correlate}), which we can re-write as
$$ \ket{0} \otimes (-1)^{f(0)}(\ket{0} - \ket{1}) + \ket{1} \otimes (-1)^{f(1)}(\ket{0} - \ket{1}) $$

\subsubsection{Hadamard 2}

In general, this step is more complicated than the single Hadamard in the Deutsch algorithm, although at least one
such transformation is always a part of it. The aim of this step is to leave the system in such a state that a
measurement of it will give the answer that we require. This is done by increasing the amplitude of the part of the
state containing the correct answer, to the detriment of other parts. Some algorithms, such as Grover's, are
probabilistic so the amplitude of the `correct' part of the state is high but not 1, and others, like the Deutsch
algorithm, are definite, so the amplitude is exactly 1. \\

In essence, what this step does is interfere all the different parts of the state in a particular way (in our
example, by a single Hadamard) to give the required answer.

\subsection{A Physical Picture}\label{pp}

So what does ESZ say is going on physically in this process?\\

To start with, it is useful to make clear the distinction in this case between branches and worlds. We will
concentrate on one branch (one `distinct macrorealm', one experimenter running the computer, etc.) and we will see
that in the computation running within this one branch we can identify many computational worlds.\\

The first set of Hadamard transformations is the easiest step to describe. We go from a state in which we can
identify a single pattern corresponding to the first register (in the state $\ket{00\ldots 0}$) to one where we can
identify $n$ patterns, each of which is a register in a different state. In our example, we start with one register
pattern in the state \ket{0} and end up with a superposition of two patterns, one of which is in the state \ket{0}
and the other in the state \ket{1}. These are now two computational worlds.\\

At this point in the proceedings there is nothing special to mark out these worlds over others that we could
identify within the same state using a different basis. What will justify our looking at these particular patterns
is how they evolve during the course of the computation, most importantly at the next step of manipulation.\\

Prior to the manipulation stage if we look at the state of the whole qubit array then we see the $n$ different
patterns of the first register, and one pattern of the second which is in its default state (in our example,
\ket{1}). During the manipulation, states of the second register ($\ket{g(\alpha)}$) become correlated with states
of the first ($\ket{\alpha}$) as the two become entangled. We can now identify multiple patterns, each of which is
two registers, which are in a state $\ket{\alpha}\otimes \ket{g(\alpha)}$. \\

Thus all the $\ket{g(\alpha)}$ states actually exist within the state --- in the common parlance of quantum
parallelism, the function $g(\alpha)$ has been evaluated for all values of $\alpha$ simultaneously. We also note
that none of the $\ket{g(\alpha)}$ are interfering with each other at this stage in the calculation. Furthermore,
we note that if our input state was not a superposition but a single (classical) value for $\alpha$, say $a$ (say
if we looked at our computer during the computation and so `collapsed' the superposition) then the state at this
point would be the single (classical) state $\{a,g(a)\} $ which we could measure unambiguously: there is no
essentially quantum feature of this step, it is purely the linearity of quantum mechanics that gives us the state
$\sum_{\alpha} \ket{\alpha}\ket{g(\alpha)}$ from the input $\ket{\alpha}$.\\

It is from this existence of all the $\ket{g(\alpha)}$, their mutual non-interference, and the knowledge of what
happens to a single value for $\alpha$ that we say that we have multiple computational worlds present, with in each
world the function $g(\alpha)$ being evaluated for a single value of $\alpha$. In the Deutsch algorithm we have two
worlds, one of which has the registers in the state $\ket{0} \otimes (-1)^{f(0)}(\ket{0} - \ket{1})$, and in the
other they are in the state $\ket{1} \otimes (-1)^{f(1)}(\ket{0} - \ket{1})$.
Of course, were we to measure the state of the second register at this point then we would only find one value of
$g(\alpha)$ --- because our measuring device would be entangled with it, and the different values of $g(\alpha)$
would each have their own associated pattern that was a measuring device measuring a certain value.\\

In the last step, what we are trying to do is extract some global property of all the values of $g(\alpha)$ that we
have in the state, and to as it were `put' that information into a world with a high amplitude, so that when we
come to measure we will be very likely to find out that information. How we do this is the key step in getting the
speed-up over classical algorithms. We can see this from two further considerations:
\begin{enumerate}
\item Entropies of preparation and measurement. The entropy of the preparation is the Shannon (classical)
information of the preparation process of a quantum state, and is bounded from below by the von Neumann entropy
(quantum) of the state thus prepared. The entropy of measurement is the Shannon information of an ensemble of
measurement outcomes, and again is bounded from below by the von Neumann entropy of the state being measured. Taken
together, these two state that the maximum amount of (classical) information that can be stored and retrieved in
one qubit is one bit. Essentially what is going on here in ESZ is that two bits are being stored in the qubit (one
in each world), but only the information content of one world (ie one bit) can be retrieved by measurement.

\item Amount of information processed in quantum computations. Steane (2000) points out that the actual
information that one gains in a quantum computation is quantitatively the same as from a classical computation of
the same length. In our example, the quantum computation gives us a 1 bit output (the answer yes/no to the question
`does $f(0)=f(1)$?') for one measurement of the system, and a similar one-measurement classical computation also
gives one bit (the value 0 or 1 of the function $f(0)$ or $f(1)$, depending on the input). Again, we can see this
clearly in an ESZ picture: the final result will always be the information content of one world. The relation
between classical and quantum computation is that in classical throughout the entire computation there is only one
computational world, from which the answer is then read at the end; in quantum there are many computational worlds
during the computation, but an observer will only ever have access to one world. From the fact that a classical
description of an $N$ qubit system requires $2^N$ bits, (and yet we can only extract $N$ bits of information from
it) we say that we have $2^N$ computational worlds. The essential difference between the two computations is that
the quantum computation worlds can interfere, and so we can manipulate the state so that information pertaining to
all worlds is included in just one world (or a few if the algorithm is probabilistic) by the end.

\end{enumerate}

In our example, this important step is done by interfering the two worlds by a Hadamard transformation so that if
the two function values are the same then we will get one result by measuring the first register, and if they are
different then we will get another. By the end of the manipulation step (\ref{manip}) we have the first register in
a different state depending on the relationship between the function values, but this information is as it were
`spread out' over both worlds, because measuring the first register at this point cannot distinguish between the
states $\ket{0} + \ket{1}$ and $\ket{0} - \ket{1}$. We therefore in this final step rotate the computational basis
so that these two states become measurably distinct as $\ket{0}$ and $\ket{1}$.\\

To summarize. The picture ESZ gives us of a quantum computation is essentially parallel classical computations the
results of which are then interfered. All of the state exists physically, and so all values of $g(\alpha)$ are
calculated and actually exist. Up until stage three of the calculation they do not interfere with each other, and
are entangled with $\ket{\alpha}$ states which also do not interfere, so we say we have $M$ computational worlds,
where $M$ is the number of distinct states. Finally, these worlds are made to interfere with one-another, and then
by measurement the experimenter becomes correlated with one of the remaining worlds, thus reading the output of the
computation.

\section{Some Objections --- and (Hopefully) Some Answers}

Debate on the utility of an Everettian picture of quantum computation has generally been carried out at an informal
level. However, some objections have also been made in the literature, most notably Steane (2000). In this section
we will deal with the issues raised in that paper in detail, and then discuss some objections that have not made it
into the literature.

\subsection{`A Quantum Computer Only Needs One Universe'}

There are two main claims made by Steane (2000). The first is that quantum computers are not, in his words, `wedded
to' Everett interpretations for their understanding (ie he is arguing against a Deutsch-style argument (Deutsch
1997) that quantum computation is proof of the existence of many worlds). The second goes further than this, and
says that the understanding of quantum computation that Everett gives is not only not necessary, but actually
unhelpful at best and misleading at worst. As we noted in the introduction, this present paper is not concerned
with the first of these two points - however the second presents a direct challenge
to the acceptability of physical pictures such as the one given above, \S\ref{pp}.\\

We first note that Steane is not dealing with ESZ in particular, but rather some fairly vague notion of a general
Everett-style theory. We will find that a lot of his arguments simply do not apply to ESZ, or have missed the mark
for anything other than a very naive Everett theory. This is not a surprising conclusion, given his remark in \S I
that he is unable to present the theory with `crystal clarity'. The main thrust of his argument is trying to show
that it is more rigorous and intuitively correct to think in terms of quantum computation happening in a single
universe rather than in `many universes'. He makes some comments, then six `remarks', then a conclusion. We will
deal with these separately.

\subsubsection{Universes}

In presenting his idea of Everett, Steane comments that the term `world/universe' seems to be used differently
depending on whether decoherence has taken place or not (in his terminology, whether the process being described in
reversible or irreversible. Of course in ESZ there is no such thing as an irreversible process). This corresponds
to our definitions of `branch' and `world'. Recall that a `branch' is a complete macrorealm, whose stability and
independent  evolution is given by decoherence, and a `world' is stable only over `relevant' timescales, and its
independence may be given by decoherence or may --- as is the case with the computational worlds above --- be given
by the process being one where the worlds do not interfere for some part.

\subsubsection{Remark 1}

The first remark was brought up in \S\ref{pp}: that the information content of the output of a quantum computation
is the same as for a classical one of the same length. Steane thus uses this to say that it is `not self-evident
that a quantum computer does exponentially more computations than a classical computer'.\\

This fact is explained, as we showed above, in a natural way from ESZ, so it is not an intuitive argument against
it. It is not an argument against the physical understanding that we get from ESZ: we can explain from the theory
just how we can get global information about a function without consecutive evaluations of it, and in doing so we
nevertheless explain why we cannot extract all the information from all worlds.

\subsubsection{Remark 2}

As well as expanding on the previous Remark, this one makes the point that often mathematical notation can be
misleading, and that there are some cases where one can interpret (naively) what has happened as many more
processes than have actually occurred. Thus there is no straight argument from the existence of decompositions of
the state such as (\ref{many}) to the existence of many worlds.\\

Steane is correct that notation can be misleading, and also that we can sometimes incorrectly say that many
processes have happened when only one has. However, there is more going on in the ESZ picture than simply
extrapolating from the existence of the decomposition (\ref{many}) to the existence of many worlds. As we have
said, what defines the worlds is their explanatory usefulness and their stability and independence. Were these
criteria not fulfilled for the $\ket{\alpha}\ket{g(\alpha)}$ states then we would have no right to call them
`worlds' in our Everett theory.\\

Furthermore, it is not only that we \emph{can} identify a worlds decomposition of the state in this case, but that
this identification does work for us explanatorily. The `extra' ontology that we are committed to by saying that
all values of $g(\alpha)$ are \emph{really} evaluated helps us explain how we can calculate global properties of
the function so quickly. The argument is not simply that we can interpret the process of a quantum computer as many
parallel evaluations, but that this interpretation does work for us by helping us explain what is going on.

\subsubsection{Remark 3}

This Remark expands on the previous one that mathematical notation can mislead, especially with reference to
(\ref{many}). Thus we cannot `read off' from the mathematics just how many computational steps have actually been
performed.\\

As above, the point is that there is more going on in ESZ than simply leaping from mathematical notation to
ontological commitment.

\subsubsection{Remark 4}

The point is made that the different worlds are not fully independent --- evolution is unitary (and, although he
does not say it, at the end of the computation the different worlds are interfered).\\

This is not a problem for ESZ as the claim is only that the worlds are independent insofar as we consider the
manipulation stage. That is all that is claimed, not that they are completely independent. The claim is that we can
\emph{identify} worlds within the state of the computation, rather than that the state is composed of many worlds.
This point is explored further below, in the discussion of Josza's objection.

\subsubsection{Remark 5}

The sensitivity to error in an $N$-qubit quantum computation is different from that of $2^N$ classical computations
running in parallel. Such a classical computer would be sensitive to errors of the order $1/2^N$, whereas from
error-correction theory we find the quantum computer to be sensitive only to O$(1/poly(N))$. Steane uses this to
question the idea that the $2^N$ calculations are actually taking place.\\

Again, this is explained naturally from ESZ. The difference with a classical parallel computer is that an error
process in a quantum computer (such as decoherence) will act on the whole state being processed. In other words, it
will act on all of the worlds identified within the state in exactly the same way. In classical parallel computing
errors can happen to individual computations (`worlds'), but because the worlds are not fully independent (see
above) in quantum computation, errors act globally. From this, we would \emph{expect} the sensitivity to error to
be O$(1/poly(N))$
--- it is not a surprise.

\subsubsection{Remark 6}

This Remark is essentially an expansion of Remark 4, that the state of the system in computation is a single
entity, not a composition of many entities. Further to this, it is a way of representing all the different states
$g(\alpha)$ without `actually' calculating them or having them really exist. An analogy is drawn between
identifying worlds in the state and calling them real, and granting reality to the individual Fourier components of
a wave.\\

The idea that the state `represents' all the $\ket{g(\alpha)}$ states without them being real is trying to show
that Everett is not the only physical explanation of quantum computation, so is not relevant to the current paper.
The analogy with Fourier components is quite interesting, however, as it is a good example to contrast with the
worlds within a state, but which looks the same mathematically.\\

We note first of all that when we are talking about physical systems (rather than mathematical idealizations), `a
wave' is itself very much a part of a high-order ontology: it is a FAPP structure, as discussed above in
\S\ref{fs}. It is the excitation of various parts of the medium (say water) in different ways at different times
--- yet we call it a single `thing' because it is relatively stable and acts independently and is explanatorily
useful (we can have a useful theory which talks about waves as single objects). We can mathematically analyse this
object in terms of its Fourier components. This is often useful mathematically; however in deciding whether or not
to grant them \emph{physical} reality, we must look at their physical usefulness. \\

The main problem with an explanation in terms of Fourier components is that they are not independent for any useful purpose.
That is, we do not usefully have transformations that act separately on each component, as we do in (\ref{many}). That is,
they do not evolve separately under the action of the transformation, so there is no explanatory gain to be had by positing
them as real. This in fact comes down to the fact that wave mechanics is not in this sense linear, whereas quantum mechanics
 is: were we to transform each Fourier component separately and then add them together, we would in general get something
 different from what we would get by transforming the sum of the components (the wave).

\subsection{In Which World did the Calculation Happen?}

One objection that is often raised to a picture of quantum computation in terms of many worlds was given by Jozsa
(2001, seminar). Although it is fairly evident that it does not impact on ESZ, it is worth dealing with in detail
to bring out one important aspect of ESZ: the breakdown in some situations of the `worlds' concept. The objection
is that by the end of the computation we cannot tell in which of the computational worlds the computer
has been --- so why do we want to say that the `worlds' are individual and separate?\\

Even asking the question in this way does not make sense in ESZ. The state of the computation contains all the
different worlds --- this is the whole point, if it had only contained one then we would not have got speed up.
This is all, remember, within a single branch, or macrorealm: within the state of the quantum computer in our
single branch we can identify many computational worlds. As we have noted many times, the worlds are not wholly
independent, and cease even for practical purposes to be independent when they interfere.\\

This last point is the important one: {\bfseries at some points during the computation we can identify worlds
within the state, and at others we cannot}. That is, the `worlds' concept, as with all emergent concepts, breaks
down at some point\footnote{A good analogy with phonons is given in Wallace (2001$b$): the concept of a phonon as
an entity is a good and useful one when they decay slowly on relevant timescales, and once the decay becomes quick
the `phonon' concept begins to break down.}. This is neither a worry nor a problem, but simply a part of the fact
that we are dealing with higher-order ontology that is only defined in certain practical situations. In a quantum
computation, we can identify worlds from after the first Hadamard until after the manipulation stage, but in the
final Hadamard transformation the worlds concept breaks down as they interfere. All we can say is that before the
transformation we could identify $p$ worlds, and after we identified $q$ worlds. The worlds do not persist
throughout the entire computation. The reason why we identify them in ESZ is that they do not need to exist
throughout in order to be useful and to count as real. Their usefulness lies in the manipulation stage, where we
can say that we have \emph{actually} calculated all the values of the function, and so we can explain what happens
in a computation and where the speed-up over classical algorithms comes from. It is also necessary for the
explanation that they do not evolve independently for the entire computation --- it is only by interfering them, in
which process the individual identities of the worlds are lost, that we can extract information pertaining to all
computed values of the function, and hence make quantum computation distinct from
classical.\\{}\\{}\\

\begin{acknowledgements}

Many thanks to Hilary Carteret, Andrew Steane, David Wallace and Vlatko Vedral for conversations on this topic, to
David Wallace for reading and commenting on \S 1, and to Vlatko Vedral for reading and commenting on the whole
paper.

\end{acknowledgements}

\end{document}